\documentclass[prb,twocolumn,superscriptaddress,showpacs]{revtex4}
\usepackage{graphicx}
\usepackage{epsfig}
\usepackage{amssymb,amsmath,amsfonts,hyperref}
\usepackage{wasysym}
\usepackage{latexsym}
\usepackage{eucal}
\usepackage{verbatim}
\usepackage[normalem]{ulem}

\usepackage{color}

\newcommand{\be}{\begin{eqnarray}}
\newcommand{\ee}{\end{eqnarray}}

\begin{document}
\title{Two-step melting of three-sublattice order in $S=1$ easy-axis triangular lattice antiferromagnets}
\author{Dariush Heidarian}
\affiliation{\small{Department of Physics, University of Toronto, Toronto, Ontario, Canada M5S 1A7}}
\author{Kedar Damle}
\affiliation{\small{Tata Institute of Fundamental Research, 1 Homi Bhabha Road, Mumbai 400005, India}}
\begin{abstract}
We consider $S=1$ triangular lattice Heisenberg antiferromagnets 
with a strong single-ion anisotropy $D$ that dominates over the
nearest-neighbour antiferromagnetic exchange $J$. In this  limit of small $J/D$, we study low temperature ($T \sim J  \ll D$) properties of such magnets by employing a low-energy description in terms of hard-core bosons with nearest neighbour repulsion $V \approx 4J + J^2/D$ and nearest neighbour
unfrustrated hopping $t \approx J^2/2D$. Using a cluster Stochastic Series Expansion (SSE) algorithm to perform sign-problem-free quantum Monte Carlo (QMC) simulations of this effective model, we establish that the ground-state three-sublattice order of the easy-axis spin-density $S^z(\vec{r})$ melts in zero field ($B=0$) in a {\em two-step} manner via an intermediate temperature phase characterized by power-law three-sublattice order with a temperature dependent
exponent $\eta(T) \in [\frac{1}{9}, \frac{1}{4}]$. For $\eta(T) < \frac{2}{9}$ in this phase, we find that  the uniform easy-axis susceptibility of an $L \times L$ sample diverges as $\chi_L \sim L^{2-9 \eta}$ at $B=0$, consistent with a recent prediction that the thermodynamic susceptibility to
a uniform field $B$ along the easy axis diverges at small $B$ as $\chi_{\rm easy-axis}(B) \sim B^{-\frac{4-18\eta}{4-9\eta}}$ in this regime.

\end{abstract}

\vskip2pc

\maketitle
\section{Introduction}
Controlled and quantitatively accurate
theoretical calculations of the low temperature properties of geometrically frustrated magnets represent a major challenge, in
part because it is difficult to devise analytical methods that
can reliably capture the competition between different low-energy
states\cite{Frustratedmagnetreview}, and in part because
most such systems are not amenable to large-scale computations using numerically exact and unbiased quantum Monte Carlo (QMC) methods.\cite{Melko_review}. The latter
difficulty has its origins in the presence of minus signs or phase factors in the statistical weights of different configurations that contribute to the partition function when
it is sampled by such methods.

Triangular lattice antiferromagnets with $S=1$ moments at each site, with
Hamiltonian
\begin{equation}
H_{\rm AF} = J\sum_{\langle \vec{r} \vec{r}' \rangle} \vec{S}_{\vec{r}} \cdot \vec{S}_{\vec{r}} - D\sum_{\vec{i}} (S^z_{\vec{r}})^2 -B\sum_{\vec{r}} S^z_{\vec{r}} \; ,
\label{eq:H_AF}
\end{equation}
in which the single-ion easy-axis anisotropy 
$D$ is much larger than the nearest-neighbour antiferromagnetic exchange $J$
between the $S=1$ moments $\vec{S}_{r}$, provide an example
where interesting physics is accessible to controlled theoretical
treatments in spite of the strongly frustrated nature of antiferromagnetic interactions. This is because the low-energy, low-temperature
physics of $H_{\rm AF}$ in the $T \sim J \ll D$ limit can be mapped~\cite{Damle_Senthil}  on to
that of hard-core bosons
on the triangular lattice with large
nearest-neighbour repulsion $V$ and parametrically small, {\em unfrustrated} boson hopping amplitude $t$. The low-energy effective Hamiltonian is thus given by:
\begin{eqnarray}
H_{\rm b} &=&
-t\sum_ {\langle \vec{r} \vec{r}'\rangle}(b^{\dagger}_{\vec{r}}b_{\vec{r}'}+ b^{\dagger}_{\vec{r}'}b_{\vec{r}} ) \nonumber \\
&&+ V\sum_ {\langle \vec{r} \vec{r}'\rangle}(n_{\vec{r}}-\frac{1}{2})(n_{\vec{r}'}-\frac{1}{2}) -\mu \sum_{\vec{r}} n_{\vec{r}} \; . \nonumber \\
&& 
\label{eq:H_b}
\end{eqnarray}
Here, $\mu $ is the chemical potential that controls deviations of the density from the half-filling, $n_{\vec{r}}=0,1$ is the boson number at site $\vec{r}$, and
$b_{\vec{r}}$ is the corresponding
boson annihilation operator. 

The crucial minus sign in front of the hopping term ($t$ is positive) renders the boson
hopping {\em unfrustrated} and ensures that matrix elements of the boson Hamiltonian $H_{\rm b}$ can, by addition of a suitable constant to Eq.~\ref{eq:H_b}, be made strictly negative in the boson occupation-number basis. This implies that one can rewrite the partition
function ${\rm Tr} e^{-\beta H_{\rm b}}$ as a sum over configurations with positive
weight in this basis. This representation can then be used to perform large-scale numerically exact quantum Monte Carlo studies\cite{Melko_review}
of the boson Hamiltonian, Eq.~\ref{eq:H_b}, on finite lattices, from which one
may draw conclusions about the low temperature physics of $H_{\rm AF}$.

In order to describe the low temperature physics of $H_{\rm AF}$ using this approach, one sets\cite{Damle_Senthil}
$\mu = 2B$, and $V \approx 4J + J^2/D$,
$t \approx J^2/2D$,
$n_{\vec{r}} = (S^z_{\vec{r}} + 1)/2$, and $b_{\vec{r}} \sim  (S^x_{\vec{r}} - iS^y_{\vec{r}})^2/2$. 
This places the corresponding boson problem in a regime in which the frustrated nearest-neighbour repulsion between hard-core bosons dominates over the unfrustrated kinetic
energy of these bosons. In this limit of $V \gg t$, the interplay of frustration and quantum
fluctuations is known to result in a ``supersolid''
ground state for $H_{\rm b}$, in which superfluidity coexists with three-sublattice (``$\sqrt{3} \times \sqrt{3}$'') density-wave order at wavevector ${\mathbf Q} \equiv (\frac{2\pi}{3},\frac{2\pi}{3})$\cite{Auerbach_Murthy,Melko_etal,Heidarian_Damle,Wessel_Troyer,Sen_etal,Heidarian_Paramekanti}.
For the easy-axis magnet, this implies\cite{Damle_Senthil}  an unusual co-existence of three-sublattice spin-density order of $S^z$ and spin nematic
order~\cite{Chen_Levy,Andreev_Grishchuk,Chandra_Coleman} for the transverse spin components ($\langle (S^{x\;,\;y})^2 \rangle \neq 0$).

Here, our interest is in the zero field ($B=0$) melting of these coexisting low-temperature orders, which translates, in boson language, to the melting of supersolid order 
at half-filling ($\mu=0$). In this boson language, which we use here and henceforth to make contact with earlier computational work on ground state properties of $H_{\rm b}$ and other closely related bosonic systems, it is clear that the superfluid order will be lost via a Kosterlitz-Thouless transition at
temperature $T_{\rm KT}$. In this article, we confirm this expectation using
detailed QMC computations that  map out this line of Kosterlitz-Thouless transitions in the $(T,V)$ phase diagram of $H_b$. 
How does three-sublattice density-wave
order melt at $\mu=0$? Two very 
different answers have been given earlier: In Ref.~\onlinecite{Boninsegni} Boninsegni and Prokofiev
have suggested that the melting proceeds even at half-filling via a
single continuous transition in the universality class of the two-dimensional three-state Potts model. Guided by the analogy\cite{Domany_Schick_Walker_Griffiths,Domany_Schick} with the two dimensional six-state
clock model\cite{Jose,Tobochnik,Challa_Landau,Rastelli_Regina_Tasi}, the present authors\cite{Heidarian_Damle}
have instead argued that the melting will proceed at half-filling via a two-step melting process characterized by an intermediate phase exhibiting power-law three-sublattice ($\sqrt{3} \times \sqrt{3}$)
order.

As was emphasized recently\cite{Damle_PRL}, both scenarios in fact
represent entirely consistent and generic possibilities for the melting of three-sublattice
order in such systems. Either possibility
can arise without fine-tuning of system parameters when the microscopic
Hamiltonian has particle-hole symmetry in boson language (which translates
to the Ising symmetry that flips the sign of the easy-axis component of the spin-density). Indeed, in the general
phase diagram of such systems (for instance, with additional further-neighbour interactions), these two possibilities represent generic behaviours that are either separated from each
other either by an intervening region of first order transitions, or connected to each
other via the multicritical point
whose universal properties were studied in Ref.~\onlinecite{Damle_PRL}.
Thus, the $\sqrt{3} \times \sqrt{3}$ ordered state  of $H_{\rm AF}$ could, in principle, display any of these possibilities upon heating. 

In this article, we study this question using QMC simulations of the effective
Hamiltonian $H_{\rm b}$, and establish that three-sublattice
order melts on heating
in a {\em two-step} manner via an intermediate phase characterized by power-law
three-sublattice order with a temperature dependent
exponent $\eta(T) \in [\frac{1}{9}, \frac{1}{4}]$. For $\eta(T) < \frac{2}{9}$ in this phase, we find that  the compressibility of an $L\times L$ sample diverges as $\kappa_L \sim L^{2-9 \eta}$ at $\mu=0$.
This implies that the uniform susceptibility of an $L \times L$ sample of the easy-axis
antiferromagnet $H_{\rm AF}$ will diverge as $\chi_L \sim L^{2-9 \eta}$ at $B=0$, consistent
with the prediction made in Ref.~\onlinecite{Damle_PRL}. By standard finite-size scaling arguments\cite{Damle_PRL}, this in turn implies  that the thermodynamic susceptibility to a uniform field $B$ along the easy axis diverges at small $B$ as $\chi_{\rm easy-axis}(B) \sim B^{-\frac{4-18\eta}{4-9\eta}}$ in the power-law three-sublattice ordered phase of $H_{\rm AF}$. As emphasized in Ref.~\onlinecite{Damle_PRL}, this provides an interesting {\em thermodynamic}
signature of the intermediate-temperature power-law ordered phase in such magnets.

The remainder of this article is organized as follows. In Sec.~\ref{ModelandMethods},
we describe the actual computational strategy used, as well as define the quantities
that are measured in our QMC simulations. In Sec.~\ref{AnalysisofResults}, we describe
the results of these simulations and analyze these using standard finite-size scaling
ideas. Finally, in Sec.~\ref{ConclusionsandOutlook}, we conclude with a
brief discussion of the potential experimental relevance of our results, and discuss
some follow-up calculations that appear to be natural extensions of the work
presented here.

\section{Model and methods}
\label{ModelandMethods}
\begin{figure}
\includegraphics[width=\columnwidth]{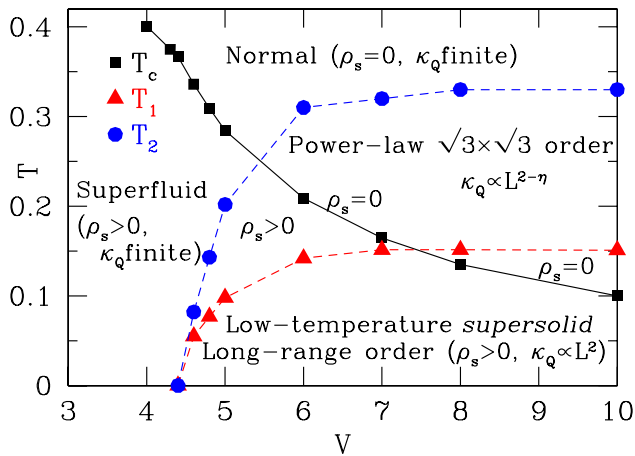}
\caption{(color online) Summary of quantum Monte Carlo results
for the phase boundaries $T_1(V)$, $T_2(V)$
and $T_c(V)$ discussed in text. Lines are guides to the eye, while points mark the measured location of the corresponding phase transition (symbol sizes have been chosen to
be indicative of the residual uncertainty in our determination of the corresponding
transition temperature, and $T$ and $V$ are measured in units of $2t$ as indicated
in the main text). }
\label{f:phase_diag}
\end{figure}

In our numerical work, we set $k_B=2t=1$ and study $H_{\rm b}$ at $\mu=0$ on $L \times L$ triangular lattices
with periodic boundary conditions, with $L$ a multiple of $6$ up to
$L=96$. Using a customized ``cluster''
implementation\cite{Heidarian_Damle,Damle_unpublished,Gros} of the Stochastic
Series Expansion (SSE)~\cite{Syljuasen0,Sandvik2,Sandvik1} QMC method
to compute equilibrium averages $\langle \dots \rangle$ at nonzero temperature $T>0$, we
focus on the finite-temperature phase diagram (Fig.~\ref{f:phase_diag}) above the $T=0$ supersolid state that obtains for $V>V_c \approx 4.6$~\cite{Auerbach_Murthy,Melko_etal,Heidarian_Damle,Wessel_Troyer,Boninsegni}.
In our cluster\cite{Heidarian_Damle,Damle_unpublished,Gros} implementation, $H_{b}$ is written as a sum $H_{b} = \sum_{\Delta,\mu} {\mathcal H}_{\Delta,\mu}$
comprising terms living on elementary triangles $\Delta$ of the triangular lattice. Here the index $\mu = 0,1,2,3$ distinguishes between the $\mu=0$ diagonal (in the number basis) term on each triangle and the three different hopping terms on the three links of each triangle. This cluster decomposition of the Hamiltonian allows
the algorithm to distinguish between minimally frustrated triangular plaquettes (with
total particle number $1$ or $2$ but not $0$ or $3$) and triangular plaquettes
that pay a high interaction energy cost by having $0$ or $3$ particles on them. This turns
out to be the key to obtaining reliable results at large values of $V$, as was demonstrated
in earlier work\cite{Heidarian_Damle}  on ground-state properties of the same model. 

We characterize superfluidity of the low-temperature
state by computing the superfluid stiffness
$\rho_s$. In order to characterize the three-sublattice
density-wave order, we focus on $n_{{\bf Q}}$, the component
of the density operator $n_{\vec{r}}$ at wavevector ${\bf Q}$
\begin{eqnarray*}
n_{{\bf Q}} &=& \sum_{\vec{r}} n_{\vec{r}}e^{i {\bf Q} \cdot \vec{r}} \; ,
\end{eqnarray*}
and measure the equal-time correlation function
$\langle m^2\rangle=\langle |n_{{\bf Q}} |^2\rangle/L^2$ as well as the corresponding fourth
moment $\langle m^4 \rangle =\langle |n_{{\bf Q}} |^4\rangle/L^4$. In addition, we measure the
static structure factor $\kappa_{\mathbf Q}= \langle |\int^{\beta}_{0} n_{{\bf Q}}(\tau)|^2\rangle/\beta L^2$, where $\beta=1/T$.  We also measure
the second moment $\overline{E_{\kappa_{\mathbf Q}}^2}$ of the SSE estimator $E_{\kappa_{\mathbf Q}}$,whose average $\overline{E_{\kappa_{\mathbf Q}}}$ over the QMC
run gives $\kappa_{\mathbf Q}$. Although $\overline{E_{\kappa_{\mathbf Q}}^2}$ is a basis-dependent
quantity, is expected to scale in the same way as
$m_{\rm st}^4 = \langle |\int_0^{\beta} n_{{\bf Q}}(\tau) |^4\rangle/\beta^2L^4$,
whose computation involves technical difficulties that we side-step
by focusing on $\overline{E_{\kappa_{\mathbf Q}}^2}$.
Finally, we also measure the   compressibility $\kappa = \langle \Delta N_{\rm tot}^2\rangle/L^2$, where $\Delta N_{\rm tot} = \sum_{\vec{r}} (n_{\vec{r}}-1/2)$.

\section{Analysis of results}
\label{AnalysisofResults}
\begin{figure}
\includegraphics[width=\columnwidth]{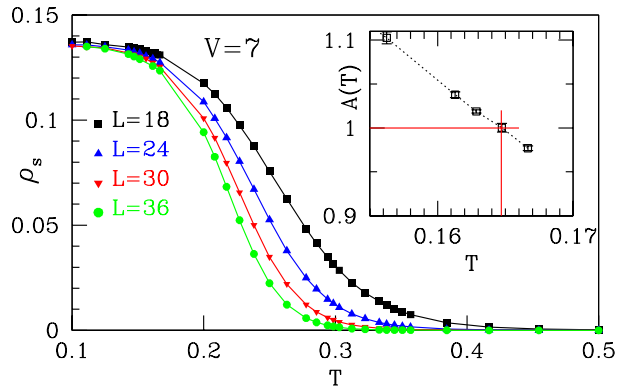}
\caption{(color online) QMC results (error bars are smaller
than symbol size) for $\rho_s(T)$ for various sizes $L$ at $V =7.0$, fit to the finite-size scaling form Eqn.~{\protect{\ref{Webereqn}}} (lines). Inset: $A(T)$ extracted from this fit provides the estimate
$T_c(V) \approx 0.164(1)$ (lines are a guide to the eye). In both the inset and the main figure, $T$ is measured in units of $2t$ as indicated
in the main text.
}
\label{f:rhosKT}
\end{figure}

We expect superfluidity to be lost on heating
via a Kosterlitz-Thouless transition at $T_c(V)$~\cite{Jose,Kosterlitz_Thouless,Nelson_Kosterlitz}.
In the vicinity of $T_c(V)$ for fixed $V$, we expect a finite-size
scaling form~\cite{Weber}
\be
\rho_s=\frac{2TA_V(T)}{\pi}\left(1+\frac{1}{2\log(L/l_V(T))}\right) \;,
\label{Webereqn}
\ee
where $l_V(T)$ and $A_V(T)$ are fitting parameters which depends on temperature.
At $T_c(V)$, we expect $A_V(T_c) =1$~\cite{Jose,Nelson_Kosterlitz,Weber}. Therefore,
to pinpoint the critical point at each $V$, we fit finite-size data for $\rho_s$
to this form and identify $T_c(V)$  with the
temperature at which the best-fit gives $A_V=1$~\cite{Harada} (see Fig.~\ref{f:rhosKT}). Using this procedure, we map out the superfluid
transition $T_c(V)$ shown in Fig~\ref{f:phase_diag}.

If three-sublattice density-wave order is lost via a single
continuous thermal phase transition at $T=T^{*}$, one expects the Binder cumulants $g_{m} = 1-\langle m^4\rangle/3\langle m^2\rangle^2$ and $g_{\kappa_{\mathbf Q}} = 1-\overline{E_{\kappa_{\mathbf Q}}^2}/3(\overline{E_{\kappa_{\mathbf Q}}})^2$ to tend to $1/3$ deep in the ordered phase, and rise monotonically to a value of $2/3$ deep in the disordered phase. Additionally, in the vicinity of the continuous transition, one expects them to collapse on to scaling functions $F_{m,\kappa_{\mathbf Q}}(\Delta T^{*})$
where $\Delta T^{*} \equiv (T-T^{*})L^{1/\nu}$, and $\nu$ is the density-wave
correlation length exponent. Such a continuous transition can therefore
be located from either Binder cumulant by looking for a crossing point of $T$-dependent curves corresponding to various sizes $L$.
On the other hand, if long-range density-wave order is lost upon heating via
a two-step melting process, with a power-law ordered intermediate
phase for $T_1 \leq T \leq T_2$, one expects $g_m$ and $g_{\kappa_{\mathbf Q}}$ to saturate at large $L$ in this power-law ordered phase to  $L$-independent limiting values
that define the universal functions $G_{m,\kappa_{\mathbf Q}}(T)$ for $T \in [T_1,T_2]$. In other words, if there is
an intermediate power-law ordered phase, we expect the
Binder cumulant curves to {\em stick} over a finite range of $T$ rather than
{\em cross} at one temperature.

\begin{figure}
\includegraphics[width=\columnwidth]{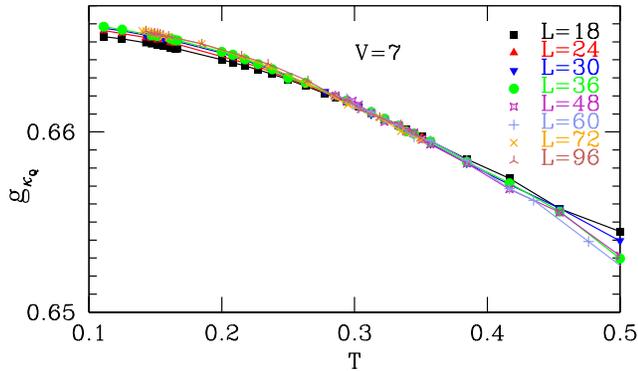}
\caption{(color online) $T$ dependence of the Binder cumulant $g_{\kappa_{\mathbf Q}}$ at $V=7$ (error bars are smaller than symbol size, and lines are a guide to the eye). Note that
the curves for different $L$ stick to each other over a finite temperature range, rather than
cross at a single point. $T$ and $V$ are measured in units of $2t$ as indicated in
the main text.}
\label{f:nqKTU7}
\end{figure}

\begin{figure}
\includegraphics[width=\columnwidth]{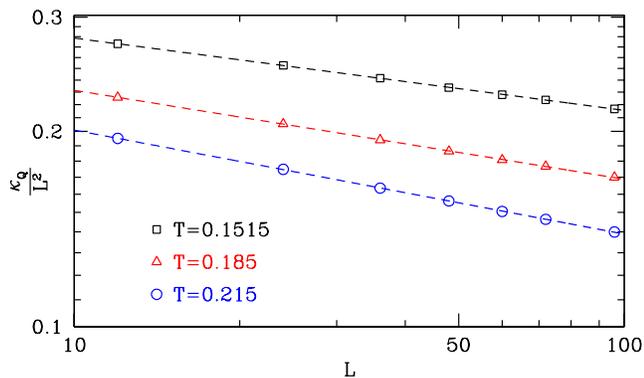}
\caption{(color online) $\kappa_{\mathbf Q}/L^2$ (error-bars are smaller than symbol size) fit to a power-law $\sim L^{-\eta(T)}$ for $T=0.1515, 0.185, 0.215$ (lines), with best-fit values $\eta(T) \approx 0.1114(2), 0.1370(3), 0.1602(3)$ respectively. }
\label{f:kappaQLogLog}
\end{figure}

With this in mind, we compare our QMC data
for $g_{\kappa_{\mathbf Q}}$ and $g_m$ with these
competing predictions for their $L$ and $T$ dependence.
We find clear evidence at all  $V>V_c$ that
the Binder cumulants {\em stick} rather than {\em cross}; a representative
example is shown in Fig~\ref{f:nqKTU7}. When the Binder cumulants
stick, we also find that $\kappa_{\mathbf Q}\sim L^{2-\eta(T)}$, with $\eta(T)$ increasing
with temperature. An illustrative example is shown in Fig.~\ref{f:kappaQLogLog}.
On the basis of this evidence, we conclude
that three-sublattice density-wave order is lost via a two-step
melting process with an intermediate temperature phase exhibiting
power-law density-wave order.

It is not entirely straightforward to obtain the precise location of 
the upper and lower phase boundaries of this power-law density-wave ordered
phase directly from the Binder cumulant data (for instance, the
example displayed in Fig~\ref{f:nqKTU7}). Therefore, we exploit
the fact that the long-distance correlations
of $\int^{\beta}_{0} n_{{\bf Q}}(\tau)$ in such a power-law ordered phase
are expected\cite{Heidarian_Damle,Domany_Schick_Walker_Griffiths,Domany_Schick,Damle_PRL} to correspond to the long-distance properties of the order-parameter field $\psi$ in the analogous power-law ordered phase of the six-state
clock model\cite{Jose,Tobochnik,Challa_Landau,Rastelli_Regina_Tasi}, 
In such six-state clock models, the long-distance correlations of $\psi$ are controlled, in renormalization group (RG)
language, by a line of fixed points~\cite{Jose},
characterized by effective free-energy $F_{\rm KT} = \int d^2r {\cal F}_{\rm KT}(\vec{r})$ with $\beta{\cal F}_{\rm KT} = \frac{1}{4 \pi g}(\nabla \theta)^2$, 
where $g \in [\frac{1}{9},\frac{1}{4}]$, and $\theta(\vec{r})$ is the phase of complex order parameter field $\psi(\vec{r})$ that is subjected
to a six-fold anisotropy. This range of $g$ corresponds
to values of $g$ for which this fixed-line is stable to vortex excitations as well as six-fold
anisotropy\cite{Jose}. For $g$ in this range, $F_{\rm KT}$ is characterized by power-law correlators $\langle \psi^{*}(\vec{r})\psi(0)\rangle \sim 1/r^{\eta(g)}$ with $\eta(g) = g$. This implies
that the power-law exponent $\eta(T)$ in such a power-law ordered phase lies
in the range $(1/9,1/4)$\cite{Jose}.

\begin{figure}
\includegraphics[width=\columnwidth]{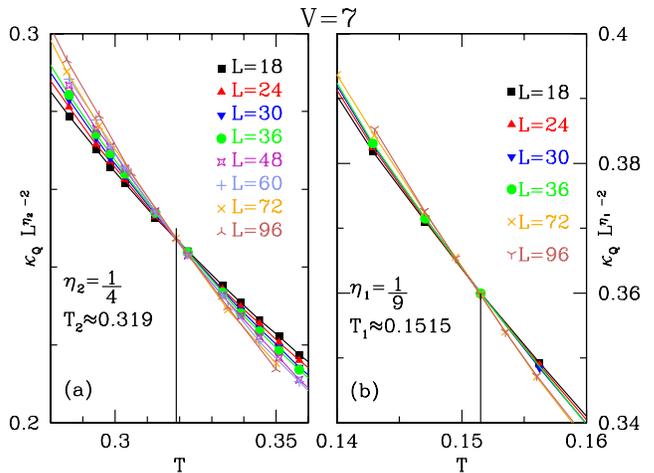}
\caption{(color online) $L^{\eta-2}\kappa_{\mathbf Q}$ (error-bars are smaller than symbol size and lines are a guide to the eye) versus temperature for various $L$
plotted in the vicinity of the upper (lower) critical point with $\eta=1/4$ ($\eta=1/9$) yield the estimates $T_2 \approx 0.319(4)$, $T_1 \approx 0.1515(5)$ for $V=7$. $T$ and $V$ are measured in units of $2t$ as indicated in the main text.}
\label{f:T2fit4}
\end{figure}

Reasoning by this analogy, we therefore expect the static structure factor $\kappa_{\mathbf Q}$ of an $L\times L$ system to obey $\kappa_{\mathbf Q}\sim L^{2-\eta(T)}$ , with $\eta(T) \in (1/9,1/4)$ in the phase with power-law
three-sublattice order. With this in mind, we determine
the upper (lower) transition
temperature $T_2(V)$  ($T_1(V)$) at each $V$ by plotting $\kappa_{\mathbf Q}L^{\frac{1}{4}-2}$ ($\kappa_{\mathbf Q}L^{\frac{1}{9}-2}$) against $T$ for various $L$ and looking for a crossing point; a representative
example is shown in Fig.~\ref{f:T2fit4}.
 Using this procedure,
we are able to map out the $V$ dependence of the phase boundaries
$T_2(V)$ and $T_1(V)$ (Fig~\ref{f:phase_diag}), and conclusively establish
a persistent intermediate phase with power-law three-sublattice
density-wave order for all $V>V_c$ studied here. For $V$ close to $V_c$,
power-law density-wave order co-exists with superfluidity, while
at larger values of $V$, this power-law ordered phase lies entirely
above the superfluid transition $T_c(V)$. Our data gives no indication that
the nature of the superfluid transition at $T_c(V)$ is affected by the presence of power-law
density-wave order, nor does it give any indication that the superfluidity
affects the nature of the power-law density-wave ordered phase or transitions
out of it. This may be understood by noting that the leading biquadratic term that couples
the square of the modulus of the superfluid order parameter to the square of the modulus
of the density-wave order parameter (in a Landau theory description of the transitions) represents an irrelevant perturbation of the decoupled critical points.

\begin{figure}
\includegraphics[width=\columnwidth]{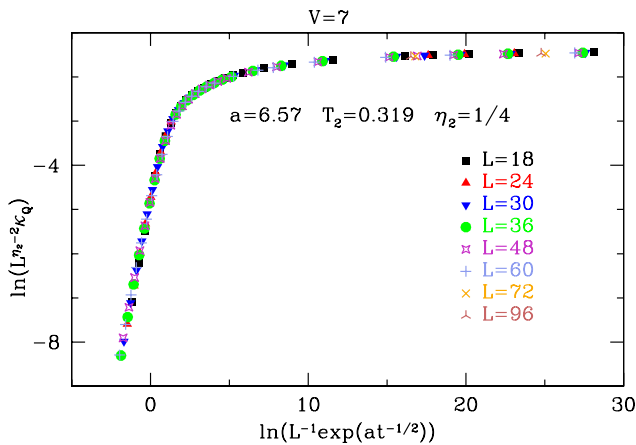}
\caption{$L$ and $T$ dependence of $\kappa_{\mathbf Q}$ for $T>T_2$ (error-bars smaller than
symbol sizes) shows scaling
collapse consistent with theoretical predictions. $T$ and $V$ are measured in units of $2t$ as indicated in the main text, and $t \equiv (T-T_2)/T_2$.}
\label{f:T2fit}
\end{figure}

 As is well-known, standard RG analysis~\cite{Jose} also predicts that 
$\langle \psi^{*}(\vec{r})\psi(0)\rangle \sim e^{-r/\xi}$ for $T>T_2$ in the six-state clock model\cite{Challa_Landau},
with $\xi \sim \exp(a|t|^{-1/2})$ where $t=(T-T_2)/T_2$ and $a$ is a dimensionless
constant. 
For the static structure factor $\kappa_{\mathbf Q}$ in a $L\times L$ system just above $T_2$, we therefore expect\cite{Challa_Landau} the analogous finite-size scaling form
$\kappa_{\mathbf Q}=L^{7/4}F_{\rm KT}(\exp(a|t|^{-1/2})/L)$
where $F_{\rm KT}$ is a universal function and $a$ is a constant.
If we use the value of $T_2(V)$ deduced from our earlier analysis, this scaling
form has a single adjustible parameter $a$ at each $V$, and the corresponding scaling
collapse of data represents a stringent test of scaling; a representative example of such
a scaling collapse
is shown in Fig.~\ref{f:T2fit}.
All of the foregoing provides clear evidence for the existence of an intermediate phase with power-law three-sublattice order at $\mu=0$, with properties consistent with RG predictions
and finite-size scaling~\cite{Jose,Weber,Challa_Landau}.

\begin{figure}
\includegraphics[width=\columnwidth]{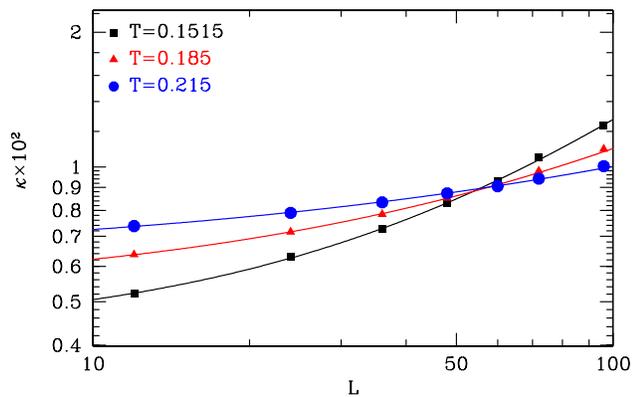}
\caption{$L$ dependence of compressibility $\kappa$ (symbols) at $V=7$ fit
to the form $a(T)+b(T)L^{2-9\eta(T)}$ (lines),
with $\eta(T)$ set equal to the best-fit value
extracted from fits of $\kappa_{\mathbf Q}$ at each temperature to the power-law form $c(T)L^{2-\eta(T)}$ in
Fig.~{\protect{\ref{f:nqKTU7}}}. $T$ and $V$ are measured in units of $2t$ as indicated in the main text.}
\label{newfig}
\end{figure}

\begin{figure}
\includegraphics[width=\columnwidth]{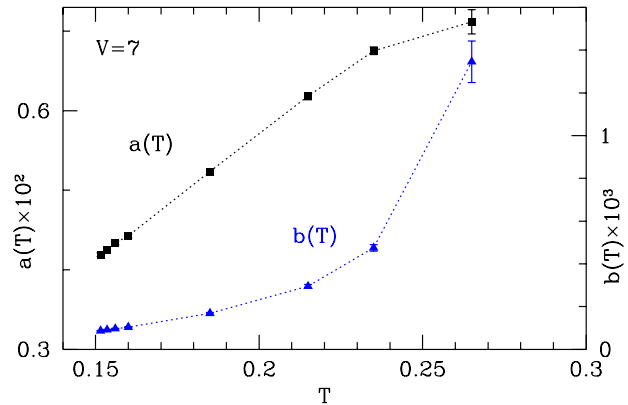}
\caption{$T$ dependence of $a(T)$ and $b(T)$ (symbols) at $V=7$. $T$ and $V$ are measured in units of $2t$ as indicated in the main text.}
\label{newfig2}
\end{figure}
The analysis of Ref.~\onlinecite{Damle_PRL} predicts that $\rho(\vec{r})$, the slowly-varying (uniform) component of the density, will also display power-law correlations in the phase with power-law three-sublattice order:
$\langle \rho(\vec{r}) \rho(0)\rangle \sim 1/r^{9\eta(T)}$ . 
For $\eta(T) < \frac{2}{9}$, these correlations decay slowly enough
that they are predicted\cite{Damle_PRL} to lead to a {\em divergent} contribution $\kappa_{\rm sing.} \sim L^{2-9\eta(T)}$ to the finite-size compressibility $\kappa_L$ of an $L \times L$ system. To check
this prediction\cite{Damle_PRL}, we fit our finite-size compressibility data in the power-law
ordered phase to the following form:
\begin{eqnarray}
\kappa_L(T) = a(T) + b(T)L^{2-9\eta(T)}
\label{nineetaform}
\end{eqnarray}
for $\eta(T) \in [\frac{1}{9},\frac{2}{9})$. In these fits, the value
of $\eta(T)$ is first obtained from a separate fit of the $L$ dependence of $\kappa_{\mathbf Q}$ to a power-law form $\sim L^{2-\eta(T)}$  at each temperature $T$, and the same
value of $\eta(T)$ is used in fitting $\kappa_L(T)$ to Eq.~\ref{nineetaform}.
An example of such
fits is shown in Fig.~\ref{newfig}, with the corresponding values of $a(T)$ and $b(T)$ displayed in Fig.~\ref{newfig2}. Based on this evidence, we
conclude that our numerical results are completely consistent
with the prediction made in Ref.~\onlinecite{Damle_PRL}.

As noted in Ref.~\onlinecite{Damle_PRL}, this result for a finite-size system
at $\mu=0$ is equivalent, via finite-size scaling, to the statement
that the thermodynamic compressibility at small $\mu$ has the singular
form: $\kappa_{\rm sing.} (\mu) \sim (\xi(\mu))^{2-9\eta(T)} \sim |\mu|^{-\frac{4-18\eta}{4-9\eta}}$.
Thus, the numerical results displayed here for
the finite-size scaling of $\kappa_L$, and the mapping between $H_{b}$ and $H_{\rm AF}$ discussed in the introduction, together establish that the power-law three-sublattice
ordered intermediate temperature phase of $H_{\rm AF}$ is characterized by a divergent
susceptibility to a uniform easy-axis field $B$: $\chi_{\rm easy-axis}(B) \sim |B|^{-\frac{4 - 18 \eta}{4-9\eta}}$ for small $|B|$. As noted in Ref.~\onlinecite{Damle_PRL}, this constitutes a striking thermodynamic signature of two-step melting of three-sublattice order in $H_{\rm AF}$, which may be of potential experimental relevance.

\section{Conclusion and Outlook}
\label{ConclusionsandOutlook}
Our results thus conclusively establish that long-range three-sublattice
order, present at low temperature in the phase diagram of $H_{\rm b}$ when
$V/2t > V_c/2t \approx 4.5$, melts in a two-step manner, via an intermediate
phase [for $T \in (T_1,T_2)$] with power-law three-sublattice order. On the other hand, superfluidity, present at low
temperature in the phase diagram of $H_{\rm b}$ for all $V/2t$, is lost
via a transition at $T_{\rm KT}$ in the Kosterlitz-Thouless universality class.
For very large values of $V/2t$, {\em i.e.} for $V/2t \gtrsim 7.5$, $T_{\rm KT}$ emerges
as the smallest of these three transition temperatures. Conversely, for $V/2t \lesssim 5.4$,
$T_{\rm KT}$ is larger than both $T_1$ and $T_2$, since these two transitions go rapidly to zero as $V$ approaches the threshold value of $V_c$ at which the ground state develops
three-sublattice order. Finally, for $5.4 \lesssim V/2t \lesssim 7.5$, the Kosterlitz-Thouless
transition occurs within the phase with power-law three-sublattice density-wave order.

Upon heating, frustrated magnets
described by $H_{\rm AF}$ with large $D/J$ (which map to $H_{\rm b}$ with $V/(2t) \approx 4D/J$) are thus expected to first undergo a Kosterlitz-Thouless transition from a low-temperature state with co-existing transverse spin-nematic order ($\langle \vec{S}_{\perp}^2\rangle \neq 0$) and longitudinal three-sublattice order ($\langle S^z({\mathbf Q}) \rangle \neq 0$) to a state with just longitudinal three-sublattice order of
the easy-axis spin density. A second
phase transition is expected to give rise to an intermediate state with power-law
three-sublattice order of the easy-axis spin density. Finally, the system is
expected to reaches a paramagnetic high-temperature state after a third transition
involving loss of power-law three-sublattice
order. In the intermediate phase with power-law three-sublattice order, the susceptibility
$\chi_{\rm easy-axis}(B)$ to a uniform field $B$ along the easy-axis is predicted to diverge as 
$|B|^{-\frac{4 - 18 \eta}{4-9\eta}}$ in the $B \rightarrow 0$ limit,
providing a {\em thermodynamic signature} of two-step melting of three-sublattice
order in such magnets.
Thus, this class of frustrated magnets is expected to exhibit several interesting
physical phenomena, and it would be interesting to identify candidate materials
whose magnetic properties are well-described by $H_{\rm AF}$.

This divergent uniform susceptibility $\chi_{\rm easy-axis}$ to
a field along the easy-axis is perhaps easier to rationalize when
the low-temperature three-sublattice ordered state is of the ferrimagnetic kind, {\em i.e.}
characterized by a net easy-axis moment that accompanies spatial symmetry breaking. This is indeed the case in the low-temperature state of $H_{\rm AF}$, as is clear upon noting that the spontaneous deviation from half-filing
identified in the low temperature state of $H_{b}$ in Ref.~\onlinecite{Heidarian_Damle} maps to a net easy-axis moment in the low-temperature state of $H_{\rm AF}$. However,
the arguments of Ref.~\onlinecite{Damle_PRL} are independent of the nature of
three-sublattice ordering, and predict that $\chi_{\rm easy-axis}$ 
would also diverge in the power-law three-sublattice ordered phase associated
with the two-step melting of {\em antiferromagnetic} three-sublattice order. A natural
and interesting follow-up to our work would therefore be to test this stronger claim in other examples of easy-axis
magnets which exhibit two-step melting of {\em antiferromagnetic} three-sublattice order.
One such example is the transverse field Ising antiferromagnet on the triangular lattice,\cite{Isakov_Moessner} in which this physics could perhaps be studied using a recently developed quantum cluster algorithm.\cite{Biswas_Rakala_Damle}

\section{Acknowledgements}
We would like to thank L.~Balents, and M.~Boninsegni and N.~Prokof'ev,  for useful correspondence regarding the results of Ref.~\onlinecite{Melko_etal} and Ref.~\onlinecite{Boninsegni} respectively.
We acknowledge use of computational resources of the Department of
Theoretical Physics of the TIFR (KD), the Universtiy of Toronto (DH), and SISSA (DH), as well as computational
resources funded by grant DST-SR/S2/RJN-25/2006 (KD). DH was the recipient of a fellowship for foreign scholars at the TIFR and a postdoctoral fellowship at SISSA (Italy) during the preliminary stages of this work. Another part of this work was completed while KD was a participant in the KITP Program on Frustrated Magnetism and Quantum Spin Liquids, funded in part by National Science Foundation grant NSF PHY11-25915. One of us would like to
acknowledge the hospitality of ICTP (Trieste) and ICTS-TIFR (Bengaluru) during the writing
of this manuscript.

\end{document}